\documentstyle[multicol,pre,aps,epsf]{revtex}
\begin{document}
\draft
\title{Slow dynamics near glass transitions in thin polymer films}
\author{Koji Fukao\cite{A} and Yoshihisa Miyamoto}
\address{
Faculty of Integrated Human Studies,
Kyoto University, Kyoto 606-8501,
Japan }

\date{Received March 23, 2001}
\maketitle

\begin{abstract}
The $\alpha$-process (segmental motion) of thin polystyrene 
films supported on glass substrate  has been investigated 
in a wider frequency range from 10$^{-3}$
Hz to 10$^4$ Hz using dielectric relaxation spectroscopy and thermal 
expansion spectroscopy. The relaxation rate of the $\alpha$-process 
increases with decreasing film thickness at a given temperature above 
the glass transition.  This increase in the relaxation rate  
with decreasing film thickness is much more enhanced near the 
glass transition temperature. The glass 
transition temperature determined as the temperature 
at which the relaxation time of the $\alpha$-process becomes a 
macroscopic time scale 
shows a distinct molecular weight dependence. It is also found that the Vogel 
temperature has the thickness dependence, {\it i.e.}, the Vogel temperature
decreases with decreasing film thickness.
The expansion coefficient of the free volume $\alpha_f$ is extracted from the
temperature dependence of the relaxation time within the free volume
theory. The fragility index $m$ is also evaluated as a function of
thickness. Both $\alpha_f$ and $m$ are found to decrease with decreasing
film thickness.
\end{abstract}

\begin{multicols}{2}
\section{\bf INTRODUCTION}

Glass transitions and related dynamics in thin polymer films 
have widely been investigated by various experimental
techniques~\cite{Confinement,Forrest1}.
One of the major motivations of such studies is to investigate 
the finite size effects of glass transitions which could be
related to the cooperative motion in supercooled liquids 
near the glass transitions. 
For this purpose, the glass transition temperature $T_{\rm g}$ 
in thin films has been measured by using several experimental 
methods such as ellipsometry~\cite{Keddie1,Keddie2}, neutron
reflectivity measurement~\cite{Wallace1}, positron annihilation life
time spectroscopy (PALS)~\cite{DeMaggio}, dielectric
measurement~\cite{Fukao1,Fukao2}.
Although contradicting experimental
results on $T_{\rm g}$ have been reported in the literatures, some common 
features seem to appear among recent experimental
results~\cite{Forrest1}.
In many cases, the glass transition temperatures in thin films are lower than
those of the bulk system if there is no strong attractive interaction 
between polymers and substrate~\cite{Keddie1,Keddie2}. In particular, 
the glass transition
temperature in freely standing films is much lower than that in 
thin polymer films supported on substrate~\cite{Forrest2,Forrest3}.

The dynamics of the $\alpha$-process (segmental motion) of polymers
in thin films have also been investigated by some techniques.
Second harmonic generation reveals that the distribution of relaxation
times broadens with decreasing film thickness, while the average
relaxation time of the $\alpha$-process remains constant for 
supported films of a random copolymer~\cite{Hall}. 
In the case of freely standing films of polystyrene, photon correlation 
spectroscopy studies indicate that the relaxation behavior of 
the $\alpha$-process in thin films is similar to that of bulk samples 
of polystyrene, except for the reduction of the $\alpha$-relaxation 
time~\cite{Forrest4}.
Because there are only a few experimental observations on 
the dynamics of thin polymer films, 
the dynamical properties of the $\alpha$-process have not yet been 
clarified in thin polymer films.

Measurements of chain mobility in thin polymer films supported on
substrate have also been done by using fluorescence recovery after 
photobleaching~\cite{Frank1} and dynamic secondary ions mass
spectroscopy~\cite{Zheng1}. A decrease in the chain mobility of the
polymer melt is observed in thin films or near the substrate by such 
experiments. This result appears to be inconsistent with the reduction
in $T_{\rm g}$ and also in the relaxation rate of segmental motions.

In previous papers~\cite{Fukao1,Fukao2}, we reported that $T_{\rm g}$ for thin
polystyrene films supported on glass substrate can be determined 
as the crossover temperature at which the temperature dependence of
the electric capacitance changes drastically
and that the dynamics of the $\alpha$-process can be determined from 
the frequency dependence of the complex dielectric constant of the films. 
We confirmed that $T_{\rm g}$ decreases with 
decreasing film thickness in the same way as observed by 
ellipsometry~\cite{Keddie1} and that the temperature $T_{\alpha}$,
at which the dielectric loss shows the peak at a given frequency of
applied electric field (from 10$^2$Hz to 10$^4$Hz), is also lower in 
thin films  than that in the bulk system, and the thickness dependence of 
$T_{\alpha}$
clearly depends on the molecular weight of the polymer.
Because the glass transition temperature is directly connected with 
the dynamics of the $\alpha$-process, the temperatures $T_{\alpha}$ and 
$T_{\rm g}$ are expected to have similarity in the properties including 
the molecular weight dependence.
However, no experimental results which support the existence 
of the molecular weight  dependence of $T_{\rm g}$ in {\it thin 
supported polymer 
films} have so far been reported~\cite{Forrest1}, in contrast with  
the case of {\it freely standing thin films}, where 
the glass transition temperature is found to strongly depend on the molecular 
weight of polymers~\cite{Forrest2,Forrest3,Dutcher1}.

In this paper, we investigate the dynamics of $\alpha$-process in thin 
films of polystyrene, especially the thickness dependence of 
the dynamics for a
wider frequency range, including lower frequencies corresponding to 
the glass transition region, 
in order to clarify the dependence of the relaxation time of the
$\alpha$-process, $\tau_{\alpha}$, and $T_{\rm g}$ on thickness and 
molecular weight. 
For this purpose, we adopt a new technique, thermal
expansion spectroscopy (TES)~\cite{Bauer1,Bauer2,Fukao3}, 
in addition to dielectric relaxation 
spectroscopy (DES). For polar materials with strong dielectric
relaxation strength, DES is a powerful methods which can cover very 
wide frequency range from 10$^{-4}$Hz to 10$^{12}$Hz by 
itself~\cite{Dixon1,Schoenhals1,Lunkenheimer1}.  Because polystyrene  
is less polar
material, however, it is difficult to obtain significant dielectric loss 
signals in lower frequency ranges by DES. In such lower frequency range,
TES is a powerful tool for non-polar materials.

On the basis of the observed relaxation rates as functions of
temperature for the thin films with various thicknesses, we obtain
the Vogel temperature and thermal expansion coefficient of the 
free volume.  The fragility index is also obtained as a function of
thickness and the non-Arrhenius behavior of the $\alpha$-process in thin 
polymer films is discussed in comparison with that in small molecules
confined in nanopores. 

\section{Experimental Details}
\subsection{Sample preparation}
Atactic polystyrene (a-PS) used in this study was 
purchased from Scientific Polymer Products, Inc. ($M_{\rm w}$=
2.8$\times$10$^5$), the Aldrich Co., Ltd. 
($M_{\rm w}$= 1.8$\times$10$^6$, 
$M_{\rm w}/M_{\rm n}$=1.03), and Polymer Source, Inc. 
($M_{\rm w}$= 6.67$\times$10$^6$, $M_{\rm w}/M_{\rm n}$=1.22),  
where $M_{\rm w}$ and 
$M_{\rm n}$ are the weight average and the number average of the
molecular weights, respectively.
Thin films of a-PS with various thicknesses were prepared on an
Al-deposited glass substrate using a 
spin-coat method from a toluene solution of a-PS. 
The thickness was controlled by changing the concentration of the solution. 
After annealing at 343K in the vacuum system for several days 
to remove solvents, Al was vacuum-deposited once more to 
serve as an upper electrode. 
Thickness of each film is evaluated from the electric capacitance
at room temperature using the equation of the flat-plate condenser
on the assumption that dielectric permittivity $\epsilon'$
does not depend on thickness, {\it i.e.}, $\epsilon'$=2.8~\cite{Yano1}.
This assumption may not appear to be a good assumption for very thin 
films. However, in our previous papers~\cite{Fukao1,Fukao2}, we showed that
the glass transition temperature $T_{\rm g}$ determined as the 
temperature at which the temperature coefficient of the electric
capacitance changes discontinuously agrees very well with that 
observed by ellipsometry even for very thin films, as mentioned in the 
Introduction. In our measurements, thickness is determined from the 
electric capacitance on the basis of the above 
assumption. These results may lead to the validity of the assumption
that $\epsilon'$ is independent of thickness for the thickness range
investigated here.

Thin films prepared according to the above procedure are located in air 
in a hand-made sample cell. The temperature is measured by an Chromel-Alumel 
thermocouple attached on the back side of glass substrate. 
Two successive thermal cycles  prior to measurements are performed 
on prepared thin films 
before the data acquisition starts in order to relax as-spun films
and obtain reproducible results. 
In the first thermal cycle the films are heated from room temperature
to 343K and then cooled down to room temperature. In the second
thermal cycle the procedure is the same as in the first one except that 
the upper-limited temperature is 378K (not 343K). 

Even after measurements up to 378K, the samples were found 
not to suffer from dewetting. However, an optical microscope measurement
showed that the surface of the upper electrode of the thin films was
slightly rough after the measurements, while that of the lower 
electrode between
the polymers and glass substrate remained smooth. This implies that 
the upper electrode
does not affect the thermal properties of thin polymer films.

\subsection{Dielectric relaxation spectroscopy}
For dielectric relaxation measurements we use an LCR meter (HP4284A)
in the range of frequency of applied electric field, $f_E$, from
20Hz to 1MHz. The measurements were performed  during heating (cooling) 
processes between room temperature
and 408K at a heating (cooling) rate of 0.5 K/min.
From the dielectric loss data in the  wide temperature and frequency 
ranges we extracted the temperature $T_{\alpha}$ at which dielectric loss
exhibits a peak at a given frequency. In order to avoid the effect 
of the extra loss peak due to the existence of a finite resistance
within the upper electrode~\cite{Fukao2,Kremer2}, we utilized the 
dielectric data up to 10kHz.

\subsection{Thermal expansion spectroscopy}
Thermal expansion spectroscopy is a new technique which has very
recently been introduced in studies on slow dynamics in supercooled liquids 
by Bauer {\it et al}~\cite{Bauer1}. In this method, a sinusoidal temperature modulation,
\begin{eqnarray}
T(t)=\langle T\rangle +T_{\omega}e^{i\omega t},
\end{eqnarray}
is applied  
to the sample, and then the corresponding change in capacitance with the same
angular frequency $\omega$ as the temperature modulation has, 
\begin{eqnarray}
C'(t)=\langle C'\rangle
- C'_{\omega}e^{i(\omega t+\delta )},
\end{eqnarray}
 is detected within a linear
response region. Here, $\omega$=2$\pi f_{\rm T}$ and $f_{\rm T}$ is the
frequency of temperature modulation, $\langle T\rangle$ is the average
temperature, $\langle C'\rangle$ is the averaged 
capacitance, $T_{\omega}$ is the amplitude of the temperature modulation
with angular frequency $\omega$, $C'_{\omega}$ is the amplitude of the
response capacitance with the angular frequency $\omega$, and $\delta$ 
is the phase lag between the temperature modulation and the 
corresponding capacitance change.  
In case of a flat-plate condenser, $C'=\epsilon'\epsilon_0S/d$, where
$\epsilon'$ is the permittivity of a-PS, $\epsilon_0$ is the
permittivity of the vacuum, $S$ is the area of the electrode 
($S=8.0$mm$^2$ in the present measurements), 
and $d$ is the thickness of the films.
If there is no dynamical process is involved, $\epsilon'\approx
\epsilon_{\infty}$, where $\epsilon_{\infty}$ is the permittivity in 
the high frequency limit, and hence the capacitance change with
temperature is directly connected with volume change in thin films via 
$\epsilon_{\infty}$, $S$ and $d$ as shown in the previous paper~\cite{Fukao2}. 

In the case of thin films of a-PS in which the area of 
the film surface remains constant with temperature change, 
the linear thermal expansion coefficient, $\alpha_{\rm n}$, normal to 
the film surface satisfies the following relation:
\begin{eqnarray}
\alpha_{\rm n} = -\zeta\frac{1}{C'(T_s)}\frac{dC'}{dT}=\zeta\frac{1}{C'(T_s)}
\frac{C'_{\omega}}{T_{\omega}}e^{i\delta},
\end{eqnarray}
where $C'(T)$ is the capacitance at temperature $T$ and $T_s$ is a
standard temperature, $\zeta\approx$0.5 for a-polystyrene~\cite{Fukao2}. 

As far as the response obeys the change in external field 
without any delay, the phase lag $\delta$ is zero. However, if there is
any dynamical process with a characteristic time of the order of
$1/\omega$ in the system, $\delta$ is no longer zero, and hence 
$\alpha_{\rm n}$ becomes a complex number. Hereafter, we denote 
$\alpha_{\rm n}$ by $\alpha_{\rm n}^*$, where
\begin{eqnarray}
\alpha^*_{\rm n}=\alpha_{\rm n}'-i \alpha_{\rm n}''.
\end{eqnarray}
Here, $\alpha_{\rm n}'$ and $\alpha''_{\rm n}$ are the real and 
imaginary parts of the complex thermal expansion coefficient.

In the
present measurements the average temperature $\langle T\rangle$ is
controlled to increase with a constant rate, 0.1 K/min or 0.5 K/min.  
The amplitude $T_{\omega}$ is set between 0.2 K and 0.6 K.
For capacitance measurements in TES the frequency of the applied electric
field was chosen to be 100kHz to avoid the interference with dielectric 
relaxations~\cite{Bauer1}.

In both DES and TES measurements, we measure (complex) capacitance of
thin films, and hence we use the same sample for the two different 
measurements. 
If $f_T=0$ and $f_E$ is
varied at a fixed temperature, we obtain the data of the frequency
dispersion of the dielectric constant. 
If $f_E$ is fixed to be a high frequency, for example, 100kHz and 
$f_T$ is varied at a given temperature, we are able to perform TES
measurement on the {\it same sample} which are used for DES
measurements. For this reason, we can obtain dynamical properties 
in a wider frequency range from 10$^{-3}$Hz to 10$^4$Hz  even
for less polar material such as a-PS by combining two powerful methods DES and 
TES.

\section{Results and Discussion}
\subsection{Dynamics of the $\alpha$-process near the glass transition 
temperature}

Figure 1 displays the temperature dependence of both the real and 
imaginary parts $\alpha'_{\rm n}$ and $\alpha''_{\rm n}$ 
measured by TES for thin 
films of a-PS with film thickness 18 nm and 247 nm and $M_{\rm w}$=
6.7$\times$10$^6$. The modulation frequencies of temperature, $f_T$, 
are 2.1mHz and 16.7mHz. The peak temperature $T_{\alpha}$ at which 
the imaginary part of $\alpha_{\rm n}^*$ has a maximum at a given
frequency $f_T$ shifts to a lower temperature with decreasing film
thickness. At the same time, the width in the temperature domain of the
transition region between the glassy state and the liquid state becomes 
broader as the thickness decreases; $\sim$15K for $d$=247nm and $\sim$30K for 
$d$=18nm. This corresponds to the fact that the distribution of the 
relaxation times
of the $\alpha$-process becomes broader with decreasing film thickness.
Our recent measurements of dielectric relaxation in thin films of 
a-PS have also revealed the broadening of the dielectric loss
peak not only in the temperature domain but also in the frequency 
domain at frequencies above 100Hz~\cite{Fukao2}. Ellipsometric studies also
support the results observed in the dielectric measurements~\cite{Kawana1}. 
The broadening observed in the present TES measurements near $T_{\rm g}$
suggests that there is a distribution in $T_{\rm g}$ within the 
thin polymer layers and that the distribution becomes broader in thinner 
films. 

The solid 
curves in Fig.1 are reproduced by combining the following two
equations. One is the Havriliak-Negami equation which can empirically
describe the frequency dependence of susceptibility 
$\alpha^*(\omega)$~\cite{HN}:
\begin{eqnarray}\label{HN}
\alpha^*(\omega)=\frac{\Delta\alpha}{(1+(i\omega\tau_{\mbox{\tiny 
HN}})^{1-\alpha_{\mbox{\tiny HN}}})^{\beta_{\mbox{\tiny HN}}}},
\end{eqnarray}
where $\omega$=2$\pi f$, $f$ is the frequency of the applied external
field, $\Delta\alpha$ is the relaxation strength, 
$\tau_{\mbox{\tiny HN}}$ is the 
relaxation time, $\alpha_{\mbox{\tiny HN}}$ and 
$\beta_{\mbox{\tiny HN}}$ are the shape 
parameters. The second one is the Vogel-Fulcher law which can 
widely be used to describe the temperature dependence of the relaxation
time $\tau_{\alpha}$ of the $\alpha$-process:
\begin{eqnarray}\label{VFT}
\tau_{\alpha}=\tau_0\exp\left(\frac{U}{T-T_0}\right),
\end{eqnarray}
where $U$ is the apparent activation energy, $T_0$ is the Vogel
temperature, and $\tau_0$ is the relaxation time at high 
temperatures~\cite{VFT1}.
In this analysis, it is assumed that the parameters $\Delta\alpha$,
$\alpha_{\mbox{\tiny HN}}$ and $\beta_{\mbox{\tiny HN}}$ in
Eq.({\ref{HN}}) are independent of temperature and that 
$\tau_{\mbox{\tiny HN}}$=$\tau_{\alpha}$.
The parameters used in Fig.1 are listed in Table I. 
Inserting the shape parameters $\alpha_{\mbox{\tiny HN}}$ and 
$\beta_{\mbox{\tiny HN}}$ into the empirical
relation involving  $\alpha_{\mbox{\tiny HN}}$, $\beta_{\mbox{\tiny HN}}$ and
$\beta_K$, $\beta_K=((1-\alpha_{\mbox{\tiny HN}})\beta_{\mbox{
\tiny HN}})^{1/1.23}$~\cite{Colmenero1},
we obtain the stretching parameter $\beta_K$ in the
relaxation function described by the stretched exponential function, 
$\phi (t)=\exp(-(t/\tau_K)^{\beta_K})$. The values of 
$\beta_K$ obtained in this manner are
$\beta_K$=0.2 and 0.3 for $d$=18nm and 247nm,
respectively. Because the smaller value of $\beta_K$ 
means
the broader distribution of the relaxation times, in this case, of the
$\alpha$-process, this result qualitatively confirms the above
experimental results obtained in the TES measurements.

\subsection{Glass transition temperature}
Figure 2 displays the $d$ dependence of $T_{\alpha}$ for a-PS films 
with three different molecular weights for the modulation frequency of
temperature $f_T$=8.3mHz. Since the relaxation time corresponding to
this frequency is of the order of $10^2$ sec, $i.e.$, a macroscopic
time scale, the temperature $T_{\alpha}$ can be regarded 
as the glass transition temperature $T_{\rm g}$.
This definition of $T_{\rm g}$ has been supported by several
experiments. For example, Hensel and Schick showed that the
calorimetric glass transition temperature determined by DSC
(differential thermal calorimetry) during linear cooling with the
rate of $-$10 K/min corresponds to the temperature at which a structural
relaxation time is about 100 sec by temperature-modulated DSC~\cite{Hensel1}. 
Bauer {\it et~al.} measured the glass transition temperatures using 
capacitive high-frequency detection in temperature ramping as well as
in harmonic temperature cycling experiments. They found that $T_{\rm g}$
determined during cooling with the rate of $-$10 K/min corresponds to 
an average structural relaxation time of about 700sec~\cite{Bauer1}.

According to the above definition we can see in Fig.2 that 
the glass transition temperature thus obtained 
decreases slightly with decreasing film thickness down to a critical 
thickness
$d_c$. Below $d_c$, $T_{\rm g}$ decreases much more rapidly with 
decreasing film thickness. 
The $d$ dependence of $T_{\rm g}$ below $d_c$ seems to be fitted by a
linear function of $d$. (Note that the horizontal axis is on a
logarithmic scale in Fig.2.)
Furthermore, the $d$ dependence of $T_{\rm g}$ clearly 
depends on the molecular weight of the polymer. As $M_{\rm w}$ 
increases, the critical 
thickness $d_c$ increases and the slope of $T_{\rm g}$ with $d$ 
below $d_c$ also changes with $M_{\rm w}$.

As listed in Table II, the values of $d_c$ changes with $M_{\rm w}$
in accordance with the radius of gyration $R_{\rm
g}$~\cite{PolymerHand1}.  This implies the existence 
of strong correlation between $d_c$ and $R_{\rm g}$. The $M_{\rm w}$ and
$d$ dependence observed in the present measurements is similar to
that of $T_{\rm g}$ observed in freely standing films of 
a-PS~\cite{Forrest2,Dutcher1} 
and also to that of $T_{\alpha}$ in supported films observed above
$T_{\rm g}$ by using dielectric relaxation spectroscopy in the
frequency range from 10$^2$Hz to 10$^4$Hz~\cite{Fukao1,Fukao2}.
From the experimental results it follows that the $M_{\rm w}$ and 
$d$ dependence can be ascribed to the confinement effects of polymer 
chains into thin film geometry. 
This effect is characteristic of the dynamics of the $\alpha$-process of 
thin polymer films and is absent in small molecules in confined geometry.

\subsection{Relaxation rate of the $\alpha$-process and Vogel 
temperature}
Figure 3 displays the dependence of the relaxation rate of the
$\alpha$-process, $f_{\rm m}$($\equiv 1/2\pi\tau_{\alpha}$, where 
$\tau_{\alpha}$ is the relaxation time), on temperature for 
thin a-PS films with various 
film thicknesses. We measured the temperature $T_{\alpha}$ at a 
given frequency $f_{\rm m}$ (=$f_E$ for DES or $f_T$ for TES) using 
the DES and TES methods and
combined the two data to obtain the relaxation rate $f_{\rm m}$ 
in a wider frequency range from 10$^{-3}$Hz to 10$^4$Hz.
The upper and lower figures are the results for 
$M_{\rm w}$=1.8$\times$10$^6$ 
and 6.7$\times$10$^6$, respectively. We can see that at a given
temperature above 390K, the relaxation rate $f_{\rm m}$ increases with 
decreasing film thickness. At a temperature near the glass transition (for
examples, 370K), the thickness dependence
of $f_{\rm m}$ is much more enhanced, $i.e.$, there is a larger 
shift in $f_{\rm m}$ near $T_{\rm g}$ than at higher temperatures
for the same amount of change in thickness. From 
this result it follows that the thickness at which the finite size effect 
begins to appear increases as the temperature approaches the 
glass transition temperature. This fact is consistent with the 
physical picture
of Adam and Gibbs that the size of the {\it cooperatively rearranging
region} (CRR) $\xi_{\rm CRR}$ increases as the temperature approaches
$T_{\rm g}$ from higher temperatures~\cite{Adam-Gibbs}, where the CRR is 
a domain in which all molecules move cooperatively. 
If the system size becomes smaller
and is comparable to $\xi_{\rm CRR}$, the dynamics should change
significantly from
those in the bulk system. It is expected that as $\xi_{\rm CRR}$ increases
the thickness at which any change in the dynamics occurs  becomes
larger. {\it The present experimental result of $f_{\rm m}\, vs.\, 1/T$ 
does qualitatively agree with the model with the growing CRR}.
This behavior has been observed in thin a-PS films with the two different 
molecular weights. 
The similar system size dependence of the relaxation rate of the 
$\alpha$-process has also been observed for the $\alpha$-process of
small molecules confined in nanopores by using dielectric relaxation
spectroscopy~\cite{Kremer1}. Such a system size dependence may be 
characteristic of the dynamics of the confined geometry.

In Fig.3 it can be seen that there is the molecular weight dependence
of the relaxation dynamics of the $\alpha$-process in thinner films. 
In order to extract the molecular weight and thickness dependence of 
$f_{\rm m}$, the observed values of $f_{\rm m}$
are fitted to the VFT equation and the parameters $U$ and $T_0$
are obtained as functions of $d$ for two different molecular weights.
We tried fitting procedures in two
different ways: 1) all the three fitting parameters $\tau_0$, $U$ and $T_0$
are adjusted to reproduce the observed values, 2) two parameters 
$U$ and $T_0$ are adjusted on condition that the parameter $\tau_0$
is fixed to be a value obtained for the bulk films.
The assumption in the case 2) corresponds to the one that $\tau_0$ is
independent of thickness, {\it i.e.}, the relaxation times at high
temperatures are the same regardless of thickness.

Figure 4 displays the thickness dependence of $T_0$ for thin films of a-PS 
with $M_{\rm w}$=1.8$\times$10$^6$ and 6.7 $\times$10$^6$. The values of 
$T_0$ in Fig.4 are obtained by the fitting procedure 2). The value of
$\tau_0$ is fixed to be 2.42$\times$10$^{-13}$ sec and 
2.87$\times$10$^{-14}$ sec for $M_{\rm w}$ 
=1.8$\times$10$^6$ and 6.7$\times$10$^6$, respectively. The values are
obtained by fitting the thickness dependence of $f_{\rm m}$ for the bulk
sample to Eq.(\ref{VFT}). In Fig.4 we can see that the Vogel temperature
decreases with decreasing thickness in the similar way to the glass 
transition temperature (Fig.2) except that there is the difference ($\sim$50K) 
in absolute values between $T_0$ and $T_{\rm g}$ at a fixed thickness.
Furthermore, we can see in Fig.4 that $T_0$ also depends on the
molecular weight of the polymers.

It should be noted that the $M_{\rm w}$ dependence of $T_0$ could not
been obtained beyond the error bars by the fitting procedure 1), 
although it is confirmed that the thickness dependence of $T_0$ could
clearly be observed, {\it i.e.}, {\it $T_0$ decreases with decreasing 
thickness whichever fitting procedure is chosen}.

\subsection{Thermal expansion coefficient of the free volume}

Figure 5 displays $1/U$ as a function of $1/d$, where the values of 
$U$ are obtained by the fitting 
procedure 2). 
In Fig.5, it is
found that $1/U$ decreases as $1/d$ increases, {\it i.e.}, 
the apparent activation energy $U$ increases with decreasing film thickness.
Here we will analyze the present result within the free volume theory.

In the free volume theory of Cohen-Turnbull~\cite{Cohen1}, the relaxation time 
of the $\alpha$-process
can be described by the following relation:
\begin{eqnarray}\label{tau_f}
\tau_{\alpha}=\tau_0\exp\left(\frac{b}{\tilde{f}}\right),
\end{eqnarray}
where $\tilde f$ is the fraction of the free volume and $b$ is a constant.
If we assume that the fraction $\tilde f$ increases linearly with temperature
as 
\begin{eqnarray}\label{f}
\tilde f=f_{\rm g}+\alpha_{f}(T-T_{\rm g}),
\end{eqnarray}
we obtain 
\begin{eqnarray}\label{tau_f2}
\tau_{\alpha}=\tau_0\exp\left(\frac{b}{\alpha_f(T-T_{\rm g})+
f_{\rm g}}\right),
\end{eqnarray}
where $\alpha_f$ is the thermal expansion coefficient of free volume
and $f_{\rm g}$ is the free volume fraction at $T_{\rm g}$.
Comparing Eq.(\ref{VFT}) with Eq.(\ref{tau_f2}), we obtain 
the Vogel temperature $T_0$ and the thermal expansion coefficient of the 
free volume $\alpha_f$ in the following way: 
\begin{eqnarray}
T_0=T_{\rm g}-\frac{f_{\rm g}}{\alpha_f}\qquad\mbox{and}\qquad\alpha_f=\frac{b}{U}.
\end{eqnarray}
Because $\alpha_f$ is proportional to $1/U$, we can see in Fig.5 that
the thermal expansion coefficient of the free volume 
$\alpha_f$ decreases with decreasing
film thickness. As shown in the straight line in Fig.5, the dependence
of $\alpha_f$ on $d$ is given as follows:
\begin{eqnarray}\label{alpha_f} 
\alpha_f=\alpha_f^{\infty}\left(1-\frac{a}{d}\right),
\end{eqnarray}
where $\alpha_f^{\infty}$ is the thermal expansion coefficient of 
the free volume in the bulk system and $a$ is a characteristic length. 
It is determined that the
values of $a$ are 4.2$\pm$0.3nm and 1.8$\pm$0.2nm for 
$M_{\rm w}$=1.8$\times$10$^6$ and 6.7$\times$10$^6$, respectively. 
This result suggests that 
within the free volume theory the mobility related to the thermal
expansion of the free volume is reduced in thinner films compared 
with that in the bulk system. 

The thermal expansion coefficient of the free volume has directly
been measured for thin films of a-PS by using PALS~\cite{DeMaggio}.
The result shows that the thermal expansion coefficient decreases
with decreasing thickness in the same manner described by
Eq.(\ref{alpha_f}). The obtained value of $a$ is 5.0$\pm$0.5nm in the case
of PALS. Furthermore, our previous measurements on temperature change in
the electric capacitance shows that the thermal expansion coefficient of 
thin films of a-PS above $T_{\rm g}$ also obeys Eq.(\ref{alpha_f}) and
$a$=2.5$\pm$0.3nm~\cite{Fukao2}. Thus, the PALS and the capacitance 
measurements on thin films
of a-PS support the validity of Eq.(\ref{alpha_f}), which is
extracted from the observed values of $f_{\rm m}$  within the
free volume theory.

Taking account of the $d$ dependence of $\alpha_f$ and $T_0$, we 
obtain the relations for
two different thicknesses $d_1$ and $d_2$,
\begin{eqnarray}
\tilde f(d_1;T)=\alpha_f(d_1)(T-T_0(d_1)),\\
\tilde f(d_2;T)=\alpha_f(d_2)(T-T_0(d_2)).
\end{eqnarray}
The present measurement shows that 
the decrease in $\alpha_f$($\sim 1/U$), 
which is associated 
with the slower relaxation, competes with the decrease in $T_0$ in thin
films, which can be regarded as the speedup of the $\alpha$-process.
Hence, we find that at a finite (crossover) temperature $T^*$ the
two straight lines corresponding to $d_1$ and $d_2$ meet with each other
in the $T$-$\tilde f$ plane.  
This result leads to the conjecture that the relaxation 
dynamics of the
$\alpha$-process in thinner films become slower than in the bulk system
above $T^*$, although the
relaxation dynamics are faster in thinner films in the temperature 
range investigated in the present measurements. If we perform the
similar measurements at still higher temperatures, we will be able to 
confirm the existence of the temperature $T^*$ at which the
$\alpha$-relaxation time in thin films becomes equal to 
that in the bulk system.
Using the present data the value of $T^*$ can be estimated: for example, 
$T^*$=431.6 K for $d_1$=12.8nm and $d_2$=247.1nm in thin films of PS
with $M_{\rm w}$=6.7$\times$10$^6$.

Tseng {\it et al.} observed fluorescence recovery after photo 
bleaching in thin polystyrene films supported on fused quartz and 
reported that the diffusion constant $D$ of dye in thin PS films falls
below the bulk values above 423K, although $D$ increases with
decreasing film thickness at a given temperature below 
423K~\cite{Tseng}.

In the previous paper~\cite{Fukao2}, we found that the existence of a 
dead layer,
where molecular mobility is highly suppressed compared with that in a
bulk-like layer, is essential to reproduce the observed thickness
dependence of the thermal expansion coefficient above $T_{\rm
g}$. Eq.(\ref{alpha_f}) suggests the existence of a similar dead layer,
which does not contribute to the thermal expansion of the free volume. 
The observed reduction in chain mobility in thin supported polymer 
films~\cite{Frank1,Zheng1} may be due to the existence of the dead
layer.

\subsection{Fragility}

As shown in the previous section, the apparent activation energy $U$ 
of the $\alpha$-process increases with decreasing film thickness.
From this result, it is expected that the fragility $m$ changes
with film thickness. Here, the fragility is a measure of non-Arrhenius
character and is defined as follows~\cite{Bohmer1}:
\begin{eqnarray}
m=\left(\frac{d\log\tau_{\alpha}(T)}{d(T_{\rm g}/T)}\right)_{T=T_{\rm g}}.
\end{eqnarray}
The observed relaxation rate of the $\alpha$-process is converted into
the relaxation time using the relation $2\pi f_{\rm m}\tau_{\alpha}=1$. 
Furthermore, the glass transition temperature $T_{\rm 
g}$ is redefined as the temperature at which the relation 
$\tau_{\alpha}(T_{\rm g})$=10$^2$sec is satisfied. Figure 6 shows the
Angell plot of the relaxation time $\tau_{\alpha}$ of the
$\alpha$-process in thin films of a-PS, where the values of
$\log_{10}\tau_{\alpha}$ are plotted as a function of $T_{\rm g}/T$. In this
figure, the fragility index is obtained as the slope of the tangential
line at $T_{\rm g}$. 

Figure 7 shows the thickness dependence of 
the fragility thus obtained for two different molecular
weights. In this figure we can see that the fragility decreases
with decreasing film thickness, {\it i.e.}, thin films of a-PS become
less fragile (more strong) as the thickness decreases.
This result means that the temperature dependence of $\tau_{\alpha}$,
which can be described by the VFT law, approaches the
simple Arrhenius law as the thickness decreases.  If it is expected that
the origin of the non-Arrhenius behavior according to the VFT law is 
related to the cooperativity in the dynamics of the $\alpha$-process,
the observed thickness dependence of fragility index leads to the
conjecture that the dynamics of the $\alpha$-process in thin films 
change from the cooperative dynamics as observed in normal liquid states
towards the single molecular dynamics.

In small molecules confined in nanopores, it is observed by dielectric 
relaxation spectroscopy that the temperature dependence of the
relaxation time of the $\alpha$-process is described by the Arrhenius
law for ethylene glycol confined in zeolite with pore size less than
0.5nm~\cite{Kremer3,Kremer4}. From this result, Kremer {\it et al.} 
estimated the minimum
number of molecules required for the emergence of non-Arrhenius
character of liquids.  The present result shows that there is the 
tendency towards the single molecular dynamics in polymers in thin films 
as well as in simple molecules confined in nanopores.

\section{Concluding remarks}

In this paper, we have investigated the dynamics of the $\alpha$-process
in thin films of a-PS in the frequency range from 10$^{-3}$Hz
to 10$^4$Hz by using dielectric relaxation spectroscopy and
thermal expansion spectroscopy. The results obtained in this study can
be summarized as follows:
\begin{enumerate}
\item
We have successfully observed the molecular weight dependence of 
the glass transition temperature for thin a-polystyrene films 
supported on glass substrate and found the existence of a crossover 
thickness at which the thickness dependence of $T_{\rm g}$ changes 
abruptly. 
\item
The crossover thickness $d_c$ strongly depends on the molecular
weight of the polymer;  the values of $d_c$ change in the
similar way to the radius of gyration of the polymer chain 
($d_c\sim R_{\rm g}$). 
\item
The relaxation time of the $\alpha$-process $\tau_{\alpha}$ decreases
with decreasing thickness. The thickness dependence of $\tau_{\alpha}$
is much more prominent near the glass transition.
\item
The Vogel temperature $T_0$ decreases with decreasing thickness.
\item
The thermal expansion coefficient of the free volume $\alpha_f$ 
decreases with decreasing thickness within the free volume theory.
\item
The fragility index $m$ decreases with decreasing thickness.
\end{enumerate}

The result 2) implies that the chain confinement effect on
the dynamics of the $\alpha$-process may be involved in the 
glass transition behavior in thin polymer
films. The molecular weight dependence of the crossover thickness 
in thin polymer films supported on glass substrate is 
qualitatively the same as that observed in freely standing films of
polymers~\cite{Forrest3}. 

In order to explain the existence of the crossover thickness 
associated strongly with $R_{\rm g}$, de Gennes proposed a model 
in which a new mode of chain motions called {\it sliding motion} 
is introduced in addition to
the segmental motion~\cite{Gennes1,Gennes2}. This new motion is a collective 
motion along the
chain, which requires a smaller free volume except for the end groups.
De Gennes shows that in the bulk state the sliding motion is suppressed
because of the end group hindrance below $T_{\rm g}$, while in thinner
films ($d<R_{\rm g}$) this motion is activated through soft surface
layers even below $T_{\rm g}$. Thus, some properties 
of $T_{\rm g}$ in the freely-standing films of polystyrene can
qualitatively be reproduced by de Gennes' model.

Although in the case of thin supported polymer films the situations are
more complicated because of the interaction between the substrate and
the polymers, the similar molecular weight dependence of $d_c$ has 
been observed as shown in the previous section. Hence, it is expected
that de Gennes' model may partly be applicable also in the present case. 
At the same time, the present form of the model can not explain 
the observed thickness and molecular weight dependence of $T_0$.
We hope that the model will be developed in full agreement with 
many aspects of the glass transition in thin polymer films.
 
\section{Acknowledgments}
The work was partly supported by a Grant-in-Aid from the Ministry 
of Education, Science, Sports and Culture of Japan.

\begin{minipage}{8.5cm}
\begin{table}\label{table1}
\caption{Parameters for the HN equation and the VFT equation to obtain
 the solid curves in Fig.1. The parameters are obtained for thin films
 of a-PS with $M_{\rm w}$=6.7$\times$10$^6$. }
\begin{tabular}{ccccccc}
$d$(nm) & $f_T$(Hz) & $\alpha_{\mbox{\tiny HN}}$ & $\beta_{\mbox{\tiny HN}}$ & 
$\beta_K$ & $U(K)$ & $T_0(K)$ \\\hline
18  & 1.67$\times$10$^{-2}$ & 0.40$\pm$0.05 & 0.22$\pm$0.03 & 0.19 & 1887 & 312.7 \\ 
18  & 2.1$\times$10$^{-3}$ & 0.38$\pm$0.07 & 0.23$\pm$0.04 & 0.20 & 1887 & 312.7 \\ 
247  & 1.67$\times$10$^{-2}$ & 0.20$\pm$0.05 & 0.29$\pm$0.03 & 0.30 & 1733 & 324.0 \\ 
247  & 2.1$\times$10$^{-3}$ & 0.23$\pm$0.06 & 0.27$\pm$0.04 & 0.27 & 1733 & 324.0 \\ 
\end{tabular}
\end{table}
\end{minipage}

\begin{minipage}{8.5cm}
\begin{table}
\caption{Critical thickness $d_c$ and radius of gyration $R_{\rm g}$ of a-PS
($M_{\rm w}$=2.8$\times$10$^5$, 1.8$\times$10$^6$ and 6.7$\times$10$^6$)}
\begin{tabular}{ccccccc}
$M_{\rm w}$ & &  2.8$\times$10$^5$ & & 
 1.8$\times$10$^6$ & &  6.7$\times$10$^6$ \\\hline
 $d_c$ (nm) & &  20 & &  34  & &  50\\ 
 $R_{\rm g}$ (nm) & &  16 & &  41 & &  79
\end{tabular}
\end{table}
\end{minipage}

\vspace*{-0.8cm}
\begin{figure}
\epsfxsize=8.5cm 
\centerline{\epsfbox{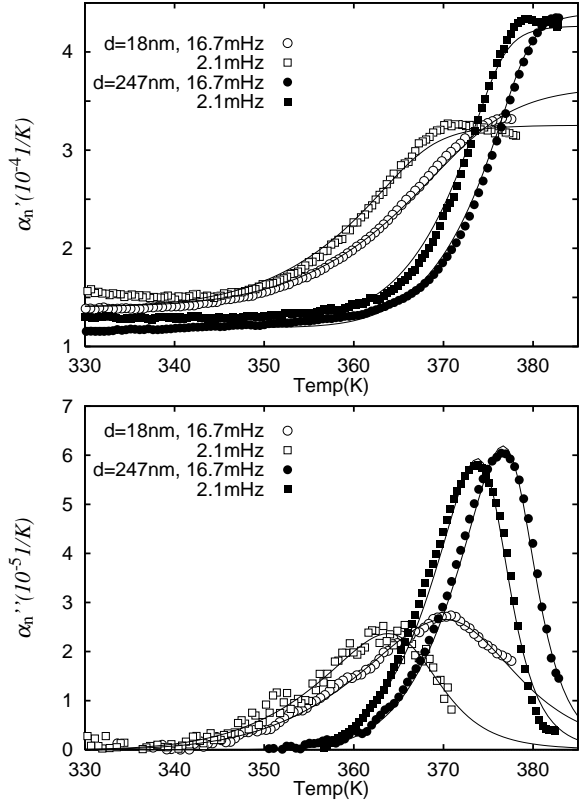}}
\vspace{-2.4cm}
\begin{minipage}{8.5cm}
\caption{
 Temperature dependence of complex linear thermal 
expansion coefficient $\alpha_{\rm n}^*$ for a-PS with film thickness 
18nm and 247nm ($f_{\rm T}$ = 16.7mHz, 2.1mHz, $M_{\rm w}$=6.7$\times$10$^6$).
The upper figure shows the real part of $\alpha_{\rm n}^*$ and the lower
one, the imaginary part. Solid lines are calculated by using the 
HN equation and the VFT equation with the parameters listed in Table I.
}
\end{minipage}
\label{fig:fig1}
\end{figure}
\vspace{0.2cm}

\vspace*{-0.8cm}
\begin{figure}
\epsfxsize=8.5cm 
\centerline{\epsfbox{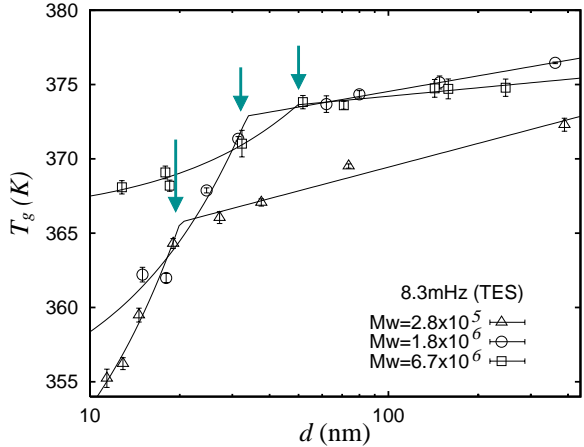}}
\vspace{0.0cm}
\begin{minipage}{8.5cm}
\caption{
Thickness dependence of the glass transition temperature
$T_{\rm g}$.  The value of $T_{\rm g}$ is measured as the temperature 
$T_{\alpha}$ at which the imaginary part of thermal 
expansion coefficients has a peak value for
the modulation frequency 8.3mHz in TES measurements.
The arrows show the crossover thickness $d_c$, below which $T_{\rm g}$
decreases abruptly with decreasing film thickness.
$M_{\rm w}$=2.8$\times$10$^5$ ($\triangle$), 1.8$\times$10$^6$ 
($\circ$) and 6.7$\times$10$^6$ ($\Box$). }
\end{minipage}
\label{fig:fig2}
\end{figure}
\vspace{0.2cm}

\vspace*{-0.8cm}
\begin{figure}
\epsfxsize=8.5cm 
\centerline{\epsfbox{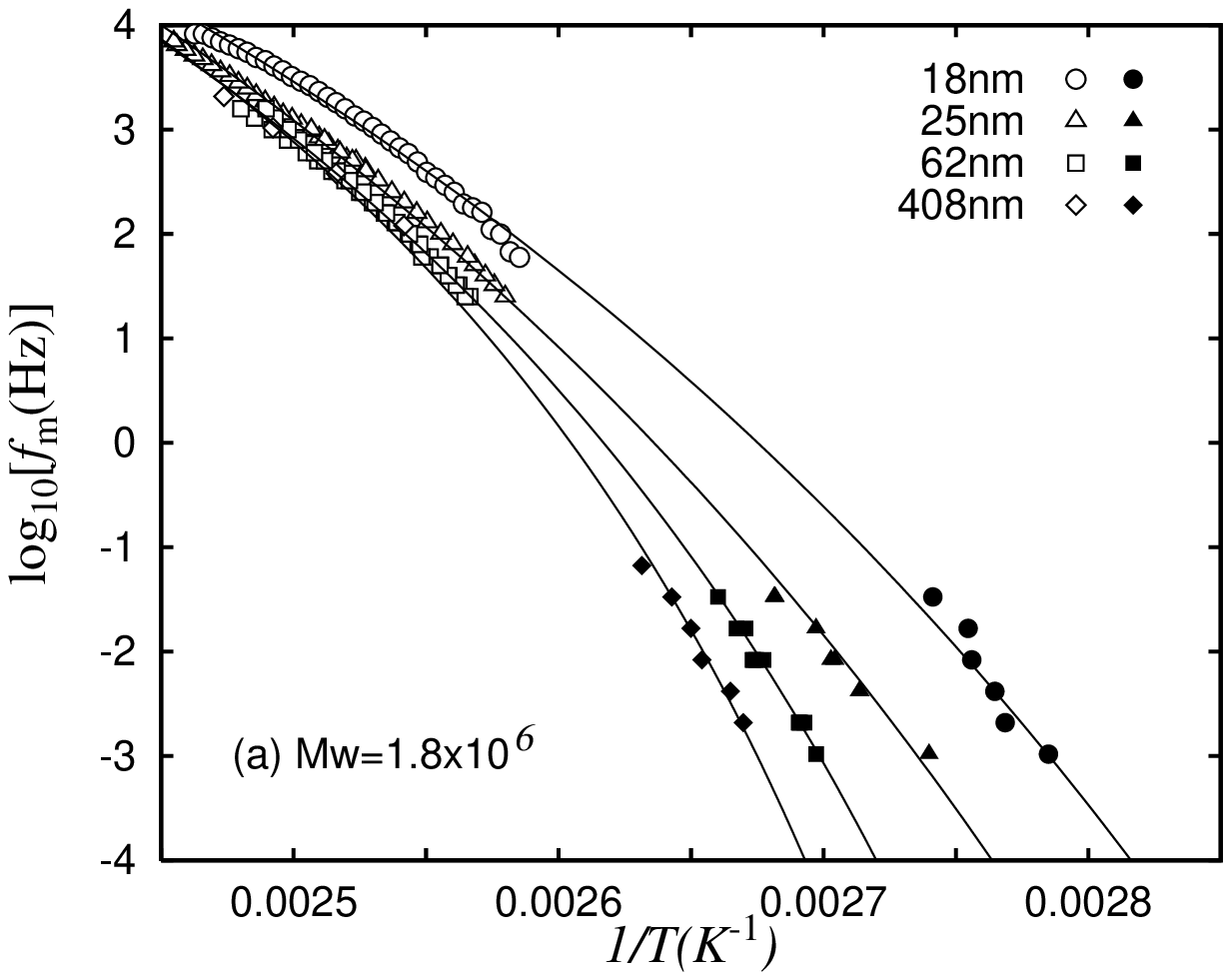}}
\epsfxsize=8.5cm 
\centerline{\epsfbox{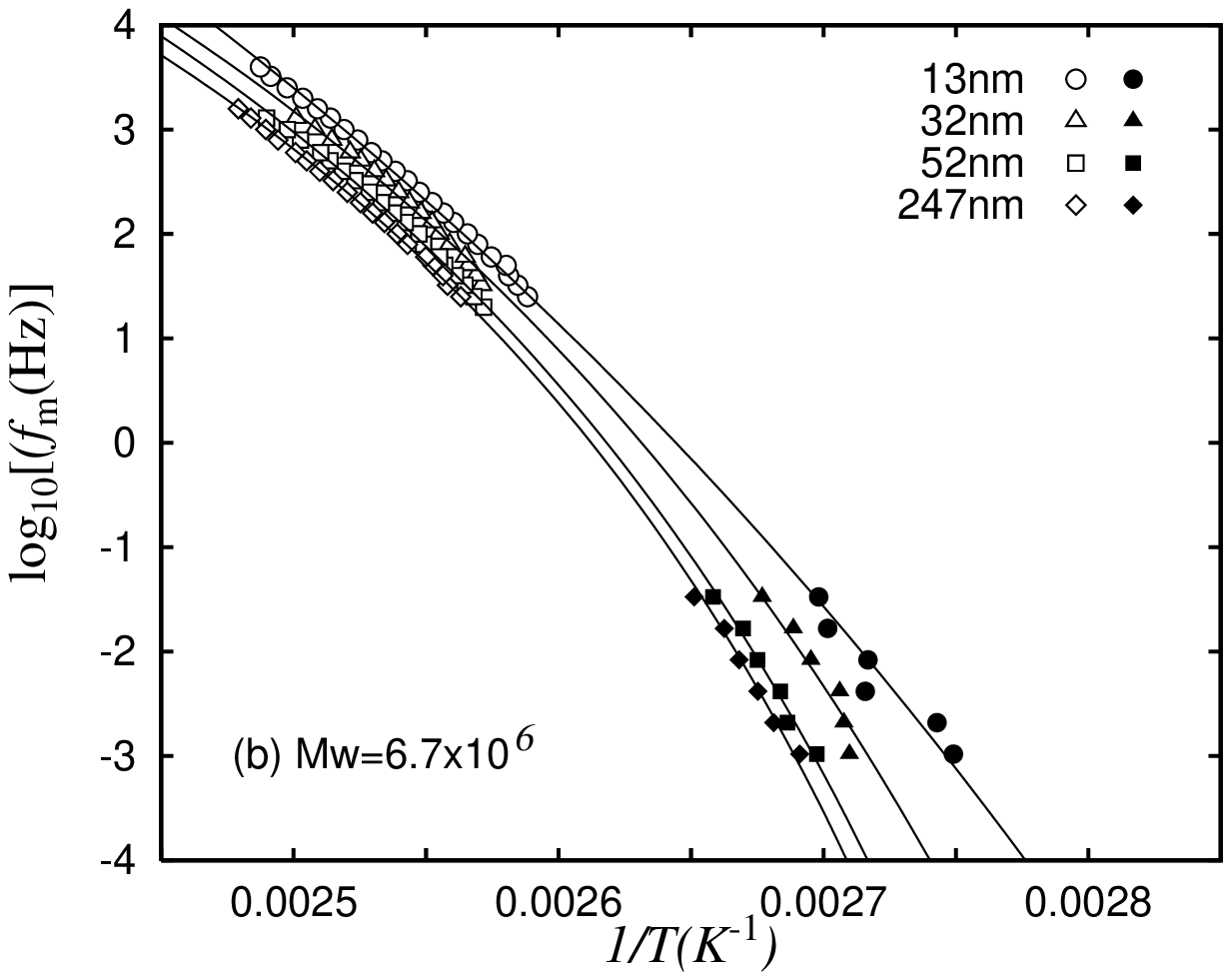}}
\vspace{0.4cm}
\begin{minipage}{8.5cm}
\caption{
Dispersion map for thin films of a-PS obtained from the peak positions
of the loss component $\alpha''_{\rm n}$ or $\epsilon''$ has the peak value 
for various film thicknesses.  The values of $f_{\rm m}$ satisfies the 
relation $2\pi f_{\rm m}\tau_{\alpha}$=1, where $\tau_{\alpha}$ is the
 relaxation time of the $\alpha$-process. Full and open symbols stand
 for the values measured by TES and DES, respectively. The thicknesses
are shown in the figure. The solid curve is calculated by using the 
VFT equation. 
(a) $M_{\rm w}$=1.8$\times$10$^6$, (b) $M_{\rm w}$=6.7$\times$10$^6$.
}
\end{minipage}
\label{fig:fig3}
\end{figure}
\vspace{0.2cm}

\newpage
\vspace*{0.0cm}
\begin{figure}
\epsfxsize=8.5cm 
\centerline{\epsfbox{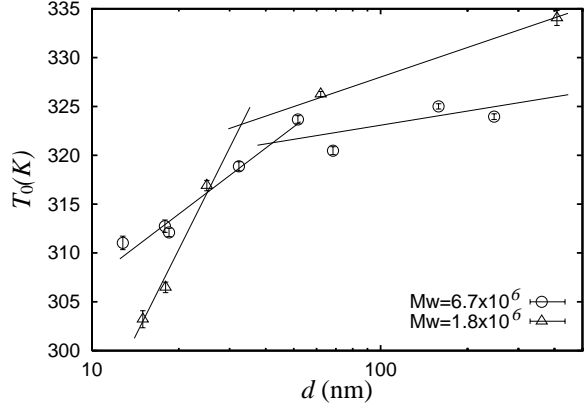}}
\vspace{0.0cm}
\begin{minipage}{8.5cm}
\caption{Thickness dependence of the Vogel temperature $T_0$ for thin
 films of a-PS ($\circ$, $M_{\rm w}$=6.7$\times$10$^6$; $\triangle$, 
$M_{\rm w}$=1.8$\times$10$^6$). The values of $T_0$ are obtained by
 fitting $f_{\rm m}\, vs. \, 1/T$ to Eq.(\protect\ref{VFT}) on condition 
 that the parameter $\tau_0$ is independent of thickness. 
The solid lines are guides for eyes. 
}
\end{minipage}
\label{fig:fig4}
\end{figure}
\vspace{0.2cm}

\vspace*{3.2cm}
\begin{figure}
\epsfxsize=8.5cm 
\centerline{\epsfbox{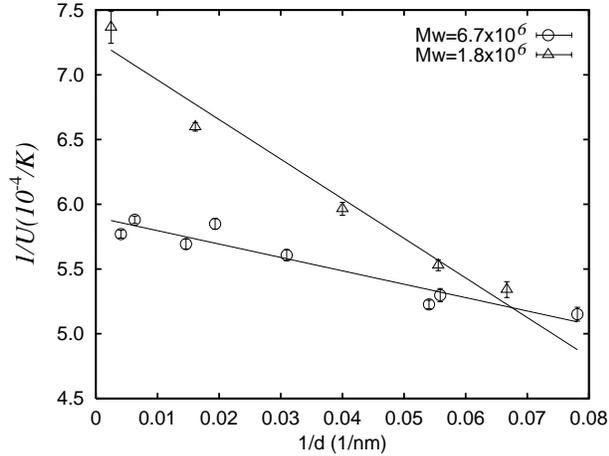}}
\vspace{0cm}
\begin{minipage}{8.5cm}
\caption{The dependence of $1/U$ on $1/d$ for thin films of a-PS 
($\circ$, $M_{\rm w}$=6.7$\times$10$^6$; $\triangle$, 
$M_{\rm w}$=1.8$\times$10$^6$). The thermal expansion coefficient of the
free volume $\alpha_f$ is proportional to $1/U$ within the free volume
 theory. The straight lines are reproduced by Eq.(\protect\ref{alpha_f}).}
\end{minipage}
\label{fig:fig5}
\end{figure}
\vspace{0.2cm}


\vspace*{0.0cm}
\begin{figure}
\epsfxsize=8.5cm 
\centerline{\epsfbox{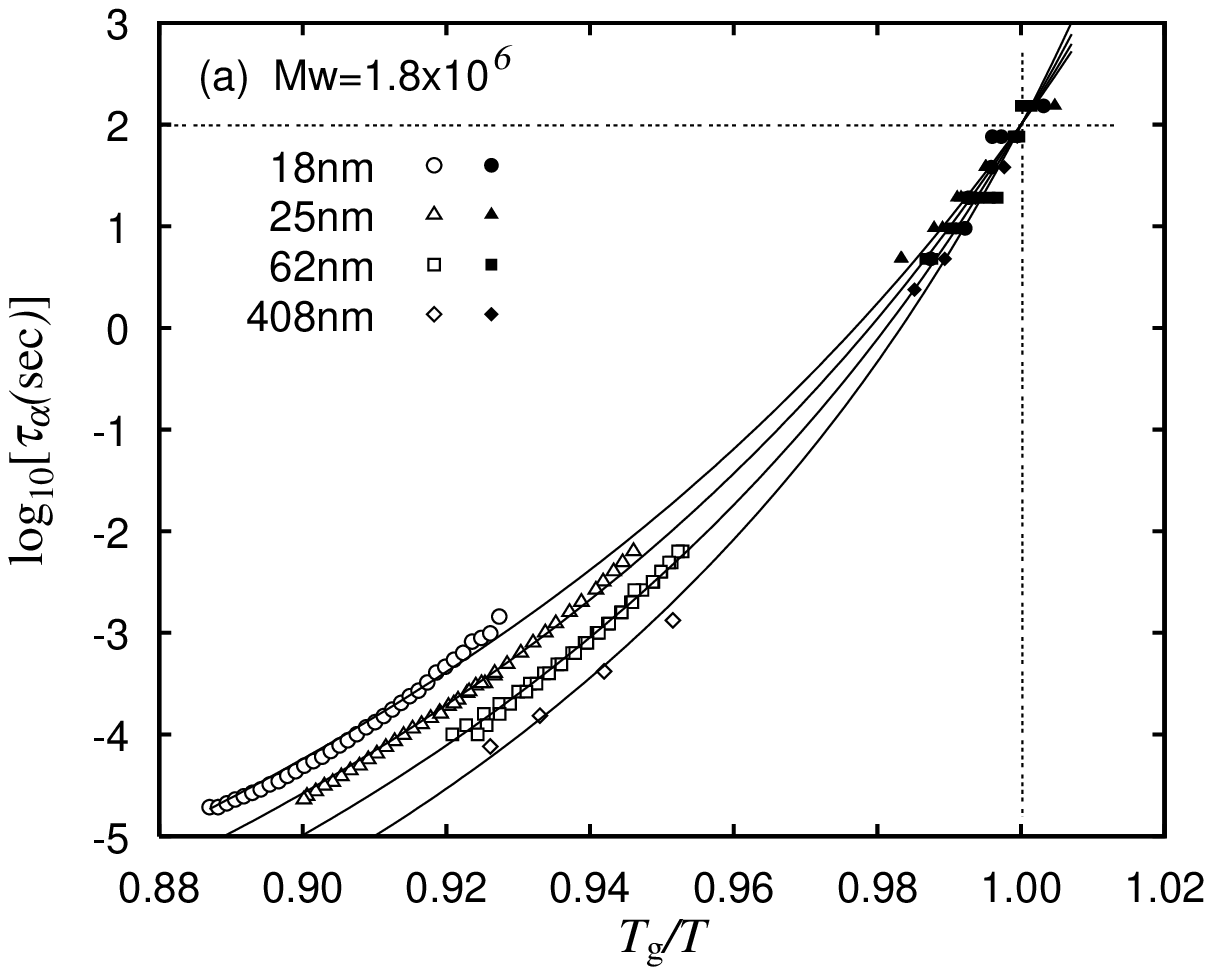}}
\epsfxsize=8.5cm 
\centerline{\epsfbox{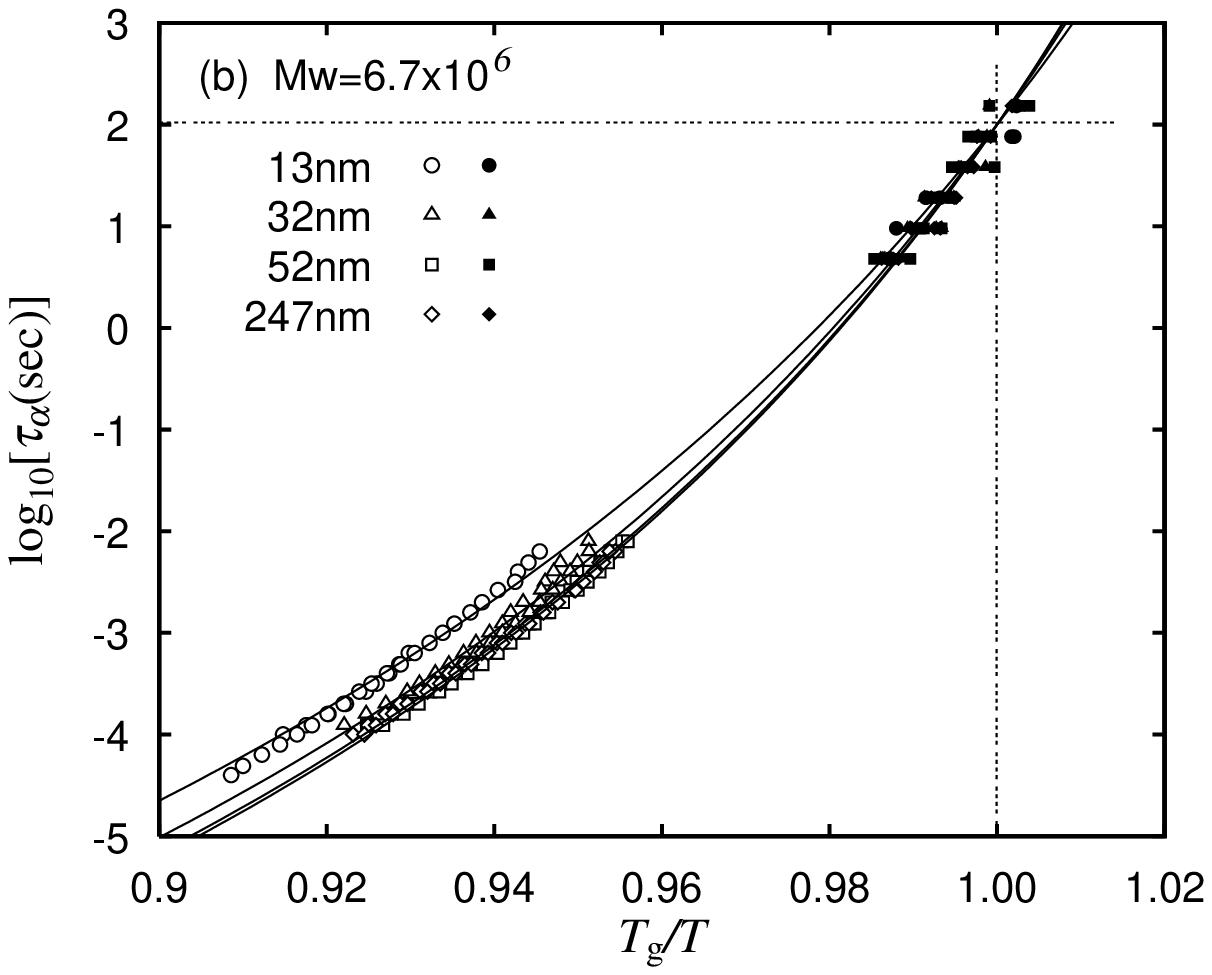}}
\vspace{0.3cm}
\begin{minipage}{8.5cm}
\caption{
The dependence of $\log_{\rm 10} \tau_{\alpha}$ on the inverse of 
the reduced 
temperature $T_{\rm g}/T$ for thin films of a-PS, 
(a) $M_{\rm w}$=1.8$\times$10$^6$, (b) $M_{\rm w}$=6.7$\times$10$^6$.
The values of $T_{\rm g}$ are obtained by the relation
 $\tau_{\alpha}(T_{\rm g})$=10$^2$sec. The symbols are the same in
 Fig.3.
}
\end{minipage}
\label{fig:fig6}
\end{figure}
\vspace{0.2cm}

\vspace*{-0.2cm}
\begin{figure}
\epsfxsize=8.8cm 
\centerline{\epsfbox{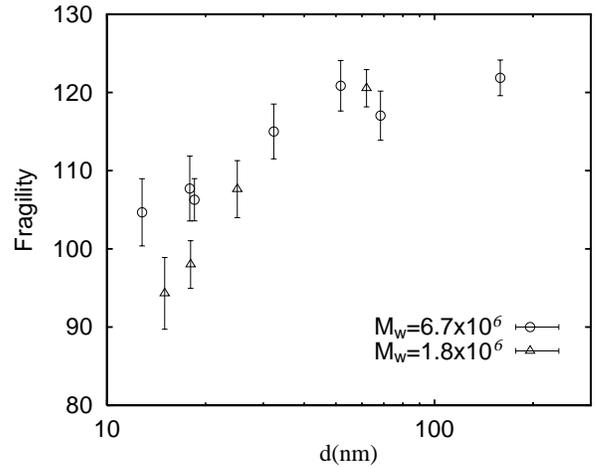}}
\vspace{0.3cm}
\begin{minipage}{8.5cm}
\caption{Thickness dependence of fragility index  $m$ for thin films of a-PS, 
$M_{\rm w}$=1.8$\times$10$^6$ ($\triangle$), $M_{\rm w}$=6.7$\times$10$^6$
($\circ$).}
\end{minipage}
\label{fig:fig7}
\end{figure}
\vspace{0.2cm}

\end{multicols}

\end{document}